\documentstyle[aps,epsf,prl,twocolumn]{revtex}
\begin{document}
\title{Formation and Dynamics of a Schr\"odinger-Cat State\\ in Continuous  
Quantum Measurement}
\author{G.P. Berman$^1$, F. Borgonovi$^{1,2}$, G. Chapline$^{1,3}$, 
S.A. Gurvitz$^{1,4}$, P.C. Hammel$^5$,\\ D.V. Pelekhov$^5$, A. Suter$^5$, 
and V.I. Tsifrinovich$^6$}
\address{$^1$Theoretical Division and CNLS,
Los Alamos National Laboratory, Los Alamos, NM 177545}
\address{$^2$Dipartimento di Matematica e Fisica, Universit\`a Cattolica,
via Musei 41 , 25121 Brescia, Italy,
and \\I.N.F.M., Gruppo Collegato  di Brescia, Italy, and I.N.F.N., sezione
di Pavia , Italy}
\address{$^3$Lawrence Livermore National Laboratory, Livermore, CA 94551}
\address{$^4$Department of Particle Physics, Weizmann Institute of Sciences, 
Rehovot 76100, Israel}
\address{$^5$MST-10, Los Alamos National Laboratory, MS K764, Los Alamos, 
NM 87545}
\address{$^6$IDS Department, Polytechnic University,
Six Metrotech Center, Brooklyn NY 11201}
\maketitle
\begin{abstract}
We consider the process of a single-spin measurement using magnetic
resonance force microscopy (MRFM) as an example of a truly continuous
measurement in quantum mechanics. This technique is also important for 
different applications, including a measurement of a qubit state in quantum
computation. The measurement takes place through the interaction of a single
spin with a quasi-classical cantilever, modeled by a quantum
oscillator in a coherent state in a quasi-classical region of
parameters. The entire system is treated rigorously within the framework of
the Schr\"odinger equation, without any artificial assumptions. Computer
simulations of the spin-cantilever dynamics, where the spin is
continuously rotated by means of {\it cyclic adiabatic inversion}, show
that the cantilever evolves into a Schr\"odinger-cat state: the
probability distribution for the cantilever position develops two
asymmetric peaks that quasi-periodically appear and vanish. For a
many-spin system our equations reduce to the classical equations of
motion, and we accurately describe conventional MRFM experiments
involving cyclic adiabatic inversion of the spin system. 
We surmise that the interaction of the cantilever with the environment 
would lead to a collapse of the
wave function; however, we show that in such a case the spin does not jump into a spin eigenstate.
\end{abstract}
{PACS:} 03.67.Lx,~03.67.-a,~76.60.-k \tolerance=10000
\section{Introduction}
Continuous quantum measurement is a very challenging problem for the
modern quantum theory \cite{1,2,3,4}. It is well-known that a
traditional measurement puts the quantum system into contact with its
environment leading to collapse of the
coherent wave function of the quantum system into one
of the eigenstates of the measurement device \cite{4}. This
phenomenon is sometimes referred to as a quantum jump. Even though traditional measurements lead eventually to wave function collapse, initially the system which is being observed may exist in a mysterious 
Schr\"odinger-cat state -- a superposition of classically distinguished states of a macroscopic system 
\cite{gerry,wan,wineland,friedman,orlando,leggett}.
If a
traditional measurement is continuously repeated, and the time
interval between two subsequent measurements approaches zero, one
faces the quantum Zeno effect: the quantum dynamics is completely
suppressed by the measurement process and the quantum system remains
in the same eigenstate of the measurement device\cite{2}. On the
other hand, quantum dynamics is not completely suppressed by a
sequence of ``non-traditional'' non-destructive measurements \cite{5,6}. These measurements assume a weak interaction 
between the quantum system and the classical measuring device. 

The other type of ``non-traditional'' non-destructive measurements are 
``truly continuous'' measurements \cite{7,7a}. A ``truly continuous'' measurement of a quantum system  means a continuous monitoring of the dynamics of a macroscopic system caused by the dynamics of a quantum system.
The result of a ``truly continuous'' measurement can be different from the output of
a sequence of many repeated measurements \cite{7}. 
Probably, the best example of a ``truly continuous'' measurement is
the very attractive idea of a single-spin measurement using magnetic
resonance force microscopy (MRFM) \cite{8}. The essence of the idea
is the following. A single spin (e.g., a nuclear spin) is placed on
the tip of a cantilever. An frequency-modulated radio-frequency
magnetic field induces a periodic change in the direction of the
spin. This can be achieved, for example, by using a well-known method
of the fast adiabatic inversion \cite{slichter}, in which the spin
follows the effective magnetic field \cite{9}. In the rotating
reference frame, this field periodically changes its
direction\cite{slichter}. The period of this motion must be much
greater than the period of precession about the effective magnetic
field, but still less than the nuclear spin-lattice relaxation time.
If this spin is placed in a strong magnetic field gradient the
rotation of the spin leads to a periodic force on the cantilever. If
this period matches the period of the cantilever
the cantilever can be driven into oscillation whose amplitude
increases to such extent that a single-spin detection may be
possible\cite{8}.

The resonant vibrations of a classical cantilever driven by a
continuous oscillations  of a single-spin $z$-component would seem to
violate the traditional expectation that coupling of a spin system to a measuring device would cause quantum jumps of the $z$-components of the spin
 \cite{10} which would prevent
successful spin detection by this method. The fundamental questions
are the following: What is the cause of quantum jumps in a 
``truly continuous'' measurement? What specific feature of the quantum
dynamics causes the collapse of the wave function of a quantum
system? The answers to these questions have both a fundamental and a
practical importance because a single-spin detection is commonly
recognized as a significant application of the MRFM, with particular
importance in the context of quantum information processing
\cite{8,9,10,11}. 

As a necessary step to approach the above problems
we  perform a detailed quantum mechanical analysis
of the coupling of a single spin to a cantilever.
 We rigorously
treat the measurement device (a quasi-classical cantilever) together
with a single spin as an isolated quantum system described by the
Schr\"odinger equation, without any additional assumptions. In
Section 2, we present the Hamiltonian and the equations of motion for
a single spin-cantilever system in the Schr\"odinger representation.
The cantilever is prepared initially in a coherent quantum state using 
parameters that
place it in a quasi-classical regime. In Section 3, we derive the
equations of motion in the Heisenberg representation; we demonstrate
that this description reduces to the classical equations of motion
for a many spin system; and we present the results of numerical
simulations of the classical spin-cantilever dynamics under the
conditions of the cyclic adiabatic inversion of the spin system. In
Section 4, we consider the quantum dynamics of the spin-cantilever
system when the spin is rotated by cyclic adiabatic inversion. Our
computer simulations explicitly demonstrate the formation of a
Schr\"odinger-cat state of the cantilever in the process of a 
``truly continuous'' quantum measurement: an asymmetric two-peak probability
distribution for the cantilever---a Schr\"odinger-cat state---is
found. This Schr\"odinger cat quasi-periodically appears and vanishes
with a period that matches that of the cyclic adiabatic inversion of
the spin. We show that the two peaks of the Schr\"odinger-cat state
each involve a superposition of both the stationary spin states. In
Section 5, we summarize our results and discuss the influence of the
environment on the dynamics of the ``truly continuous'' quantum measurement. In
particular, we argue that after the collapse of the wave function the
spin does not jump into one of its stationary states.

\section{The Hamiltonian and the equations of motion}

We consider the cantilever--spin system shown in Fig. 1.
A single spin ($S=1/2$) is placed on the cantilever tip. The tip can
oscillate only in the $z$-direction. The ferromagnetic particle,
whose magnetic moment points in the positive $z$-direction, produces
a non-uniform magnetic field at the spin.
\begin{figure}
\epsfxsize 8cm
\epsfbox{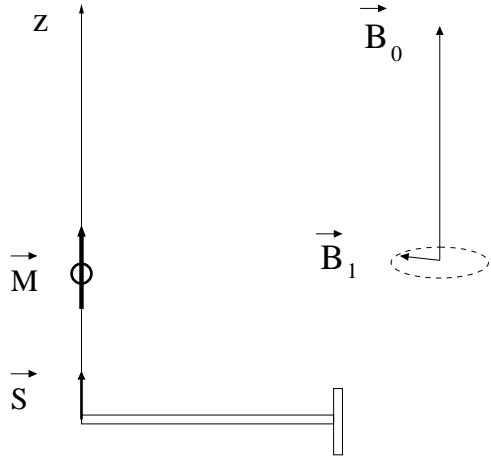}
\narrowtext
\caption{
A schematic setup of the cantilever-spin system.
$\vec{B}_0$ is the uniform permanent magnetic field; $\vec{B}_1$ is the
rotating magnetic field; $\vec{S}$ is a single spin ($S=1/2$); $\vec{ M}$ 
is the magnetic moment of the ferromagnetic particle.}
\label{canti}
\end{figure}
The uniform magnetic field, $\vec B_0$, oriented in the positive
$z$-direction, determines the ground state of the spin. The rotating
magnetic field, $\vec B_1$, induces transitions between the ground
and the excited states of the spin. The origin is chosen to be the
equilibrium position of the cantilever tip with no ferromagnetic
particle. The rotating magnetic field can be represented as,
$$
B_x=B_1\cos(\omega t+\varphi(t)),~B_y=-B_1\sin(\omega t+\varphi(t)),\eqno(1)
$$
where $\varphi(t)$ describes a smooth change in phase required for a
cyclic adiabatic inversion of the spin ($d\varphi/dt\ll\omega$).

In the reference frame rotating with $\vec B_1$, the Hamiltonian of
the system is,
$$
{\cal H}={{P^2_z}\over{2m^*_c}}+{{m^*_c\omega_c^2Z^2}\over{2}}-
\hbar\Bigg(\omega_L-\omega-{{d\varphi}\over{dt}}\Bigg)S_z-\eqno(2)
$$
$$
\hbar\omega_1S_x-g\mu{{\partial B_z}\over{\partial Z}}ZS_z.
$$
In Eq. (2), $Z$ is the coordinate of the oscillator which describes
the dynamics of the quasi-classical cantilever tip; $P_z$ is its
momentum, $m^*_c$ and $\omega_c$ are the effective mass and the
frequency of the cantilever; $S_z$ and $S_x$ are the $z$- and the
$x$-components of the spin; $\omega_L$ is its Larmor frequency;
$\omega_1$ is the Rabi frequency (the frequency of the spin
precession about the field $B_1$ at the resonance condition:
$\omega=\omega_L$, $\dot\varphi=0$); $g$ and $\mu$ are the $g$-factor
and the magnetic moment of the spin. The parameters in (2) can be
expressed in terms of the magnetic field and the cantilever
parameters:
$$
m^*_c=m_c/4,~\omega_c=(k_c/m^*_c)^{1/2},~\omega_L=\gamma B_z,~\omega_1=
\gamma B_1,\eqno(3)
$$
where $m_c$ and $k_c$ are the mass and the force constant of the
cantilever; $B_z$ includes the uniform magnetic field, $B_0$, and the
magnetic field produced by the ferromagnetic particle;
$\gamma=g\mu/\hbar$ is the gyromagnetic ratio of the spin.

Next, we introduce the following ``quants'' of the oscillator
(cantilever): energy ($E_c$), force ($F_c$), amplitude ($Z_c$), and
momentum ($P_c$),
$$
E_c=\hbar\omega_c,~F_c=\sqrt{k_cE_c},~Z_c=\sqrt{E_c/k_c},~P_c=\hbar/Z_c.
\eqno(4)
$$
Using these quantities and setting $\omega=\omega_L$, we rewrite the
Hamiltonian (2) in the dimensionless form,
$$
{\cal H}^\prime={\cal H}/\hbar\omega_c=(p^2_z+z^2)/2+
\dot\varphi S_z-\varepsilon S_x-2\eta zS_z,\eqno(5)
$$
where,
$$
p_z=P_z/P_c,~z=Z/Z_c,~\varepsilon=\omega_1/\omega_c,~\dot\varphi=d\varphi/
d\tau,~\tau=\omega_ct,
\eqno(6)
$$
$$
\eta=g\mu(\partial B_z/\partial Z)/2F_c.
$$
To estimate the ``quants'' in (4) and the dimensionless parameters in
(5), we use parameters from the MRFM measurement \cite{9} of protons
in ammonium nitrate,
$$
 \omega_c/2\pi=1.4\times 10^3 \text{ Hz}
 ,~k_c=10^{-3}\text{ N/m},~B_1=1.2\times 10^{-3}\text{ T,}\eqno(7)
$$
$$
 \partial B_z/\partial Z=600\,\text{ T/m,} ~\gamma/2\pi=4.3\times 10^7 \, 
\text {Hz/T.}
$$
Using these values, we obtain,
$$
 E_c=9.2\times 10^{-31}\text{ J},~F_c=3\times 10^{-17} \text{ N}
 ,~Z_c=3\times 10^{-14}\text{ m},\eqno(8)
$$
$$
 P_c=3.5\times 10^{-21}\text{ kgm/s},~\varepsilon=37, ~\eta=2.8\times 10^{-7}.
$$
The dimensionless Schr\"odinger equation can be written in the form,
$$
i\dot\Psi={\cal H}^\prime\Psi,\eqno(9)
$$
where,
$$
\Psi(z,\tau)=\left(\matrix{\Psi_1(z,\tau)\cr
\Psi_2(z,\tau)\cr}\right),\eqno(10)
$$
is a dimensionless spinor, and $\dot\Psi=\partial\Psi/\partial\tau$. Next, we
expand the functions, $\Psi_1(z,\tau)$ and $\Psi_2(z,\tau)$, in terms
of the eigenfunctions, $|n\rangle$, of the unperturbed oscillator
Hamiltonian,
$(p^2_z+z^2)/2$,
$$
 \Psi_1(z,\tau)=\sum_{n=0}^\infty A_n(\tau)|n\rangle,~\Psi_2(z,\tau)
 =\sum_{n=0}^\infty B_n(\tau)|n\rangle,\eqno(11)
$$
$$
  |n\rangle=\pi^{1/4}2^{n/2}(n!)^{1/2}e^{-z^2/2}H_n(z),
$$
where $H_n(z)$ is the Hermitian polynomial. Substituting (10) and
(11) in (9), we derive the coupled system of equations for the
complex amplitudes, $A_n(\tau)$, and $B_n(\tau)$,
$$
i\dot A_n=(n+1/2+\dot\varphi/2)A_n-\eqno(12)
$$
$$
(\eta/\sqrt{2})(\sqrt{n}A_{n-1}+\sqrt{n+1}A_{n+1})-(\varepsilon/2)B_n,
$$
$$
i\dot B_n=(n+1/2+\dot\varphi/2)B_n+
$$
$$
(\eta/\sqrt{2})(\sqrt{n}B_{n-1}+\sqrt{n+1}B_{n+1})-(\varepsilon/2)A_n.
$$
To derive Eqs (12), we used the well-known expressions for creation and 
annihilation operators,
$$
a|n\rangle=\sqrt{n}|n-1\rangle,~a^\dagger|n\rangle=\sqrt{n+1}|n+1\rangle,
\eqno(13)
$$
$$
[(p^2_z+z^2)/2]|n\rangle=(n+1/2)|n\rangle,
$$
$$
z=(a^\dagger+a)/\sqrt{2},~ p_z=i(a^\dagger-a)/\sqrt{2},~[a,a^\dagger]=1.
$$
\section{The Heisenberg representation and classical equations of motion}
In this section, we will find the relation between the Schr\"odinger equation 
and the classical equations of motion for the spin-cantilever system. For
this, we present the operator equations of motion in the Heisenberg
representation, 
$$
\dot z(\tau)=p_z(\tau),\eqno(14)
$$
$$
\dot p_z(\tau)=-z(\tau)+2\eta S_z(\tau),
$$
$$
\dot S_x(\tau)=-\dot\varphi S_y(\tau)+2\eta z(\tau)S_y(\tau),
$$
$$
\dot S_y(\tau )=\dot\varphi S_x(\tau)+\varepsilon S_z(\tau)-2\eta z(\tau)S_x(\tau),
$$
$$
\dot S_z(\tau)=-\varepsilon S_y(\tau).
$$
In Eqs (14), the time-dependent Heisenberg operators are related to the 
Schr\"odinger operators in the standard way, e.g.,
$z(\tau)=U^\dagger(\tau)zU(\tau)$, where $U(\tau)=\hat
T\exp(-i\int_0^\tau{\cal H}^\prime(\tau^\prime)d\tau^\prime)$, where $\hat T$
is the time-ordering operator.

To derive classical equations we first present the equations of motion for 
averages of the Heisenberg operators. We have from Eqs (14),
$$
\langle \dot z(\tau)\rangle=\langle p_z(\tau)\rangle,\eqno(15)
$$
$$
\langle\dot p_z(\tau)\rangle=-\langle z(\tau)\rangle+2\eta \langle S_z(\tau)
\rangle,
$$
$$
\langle \dot S_x(\tau)\rangle=-\dot\varphi \langle S_y(\tau)\rangle+2\eta 
\langle z(\tau)\rangle\langle S_y(\tau)\rangle+2\eta R_1,
$$
$$
\langle \dot S_y(\tau)\rangle=\dot\varphi \langle S_x(\tau)\rangle+\varepsilon 
\langle S_z(\tau)\rangle-2\eta \langle z(\tau)\rangle \langle
S_x(\tau)\rangle-2\eta R_2 , 
$$
$$
\langle \dot S_z(\tau)\rangle=-\varepsilon \langle S_y(\tau)\rangle,
$$
where $R_1$ and $R_2$ are quantum correlation functions,
$$
R_1=\langle zS_y\rangle-\langle z\rangle
\langle S_y\rangle, ~R_2=\langle zS_x\rangle-\langle z\rangle
\langle S_x\rangle.\eqno(16)
$$

Now consider $N$ spins interacting with a cantilever. At $\tau=0$, some of 
these spins are in their ground states, and others are in their excited
states. We introduce a dimensionless ``thermal'' magnetic moment,
$\sum_k\langle \vec S_k\rangle$. Neglecting quantum correlations under the
conditions, %
$$ |\langle z\sum_kS_{x,y}\rangle-\langle z\rangle\sum_k
\langle S_{x,y}\rangle|\ll |\langle z\rangle\langle 
\sum_k S_{x,y}\rangle|,\eqno(17)
$$
we derive the classical equations for the spin-cantilever system,
$$
\dot z=p_z,\eqno(18)
$$
$$
\dot{p_z}=- z +2\eta\Delta N S_z,
$$
$$
\dot{ {\vec S}}=[ {\vec S}\times {\vec b_e}],
$$
where the angular brackets are omitted; $\vec b_e$ is the effective 
dimensionless magnetic field with components,
$$
b_{ex}=\varepsilon,~b_{ez}=-\dot\varphi+2\eta z.\eqno(19)
$$
The thermal dimensionless magnetic moment, $\sum_k\langle {\vec S}_k \rangle$, 
is represented as $\Delta N\langle  S\rangle $, where $\Delta N$ is the
difference in the population of the ground state and the excited state of the
spin system (i.e. the effective number of spins at given temperature) at time
$\tau=0$.

The second term in the expression for $b_{ez}$ describes the nonlinear effects 
in the dynamics of the classical magnetic moment. In terms of the dimensional
quantities, $Z$, $P_z$, and $\vec {\cal M}=\gamma\hbar \Delta N\vec S$, the
classical equations of motion have the form, %
$$
{{dZ}\over{d t}}={{P_z}\over{m^*_c}},\eqno(20)
$$
$$
{{dP_z}\over{dt}}=-k_cZ+{\cal M}_z{{dB_z}\over{dZ}},
$$
$$
{{d{\vec {\cal M}}}\over{dt}}=[\vec {\cal M}\times\vec B_e],
$$
$$
B_{ex}=B_1,~B_{ez}=-{{1}\over{\gamma}}{{d\varphi}\over{dt}}+{{\partial B_z}
\over{\partial Z}}Z.
$$

To estimate the amplitude of the cantilever vibrations for the experimental 
parameters (7), we set $S_z=(1/2)\cos \tau$ in (18). Then, the driven
oscillations of the cantilever are given by the expression, %
$$
z={{1}\over{2}}\Delta N\eta\tau \sin\tau.\eqno(21)
$$
Next, to estimate the amplitude of the stationary vibrations of the
cantilever within the Hamiltonian approach, we put $\tau=Q_c$, where
$Q_c$ is the quality factor of the cantilever. (The value $\tau=Q_c$
corresponds to time $t=t_c$, where $t_c=Q_c/\omega_c$ is the the time
constant of the cantilever.) Taking parameters from the experiment
\cite{9},
$$
\Delta N=2.9\times 10^9,~Q_c\approx 10^3,\eqno(22)
$$
we obtain for the stationary amplitude of the cantilever,
$$
z=\Delta N\eta Q_c\approx 8.1\times 10^5.
$$
\begin{figure}
\epsfxsize 8cm \epsfbox{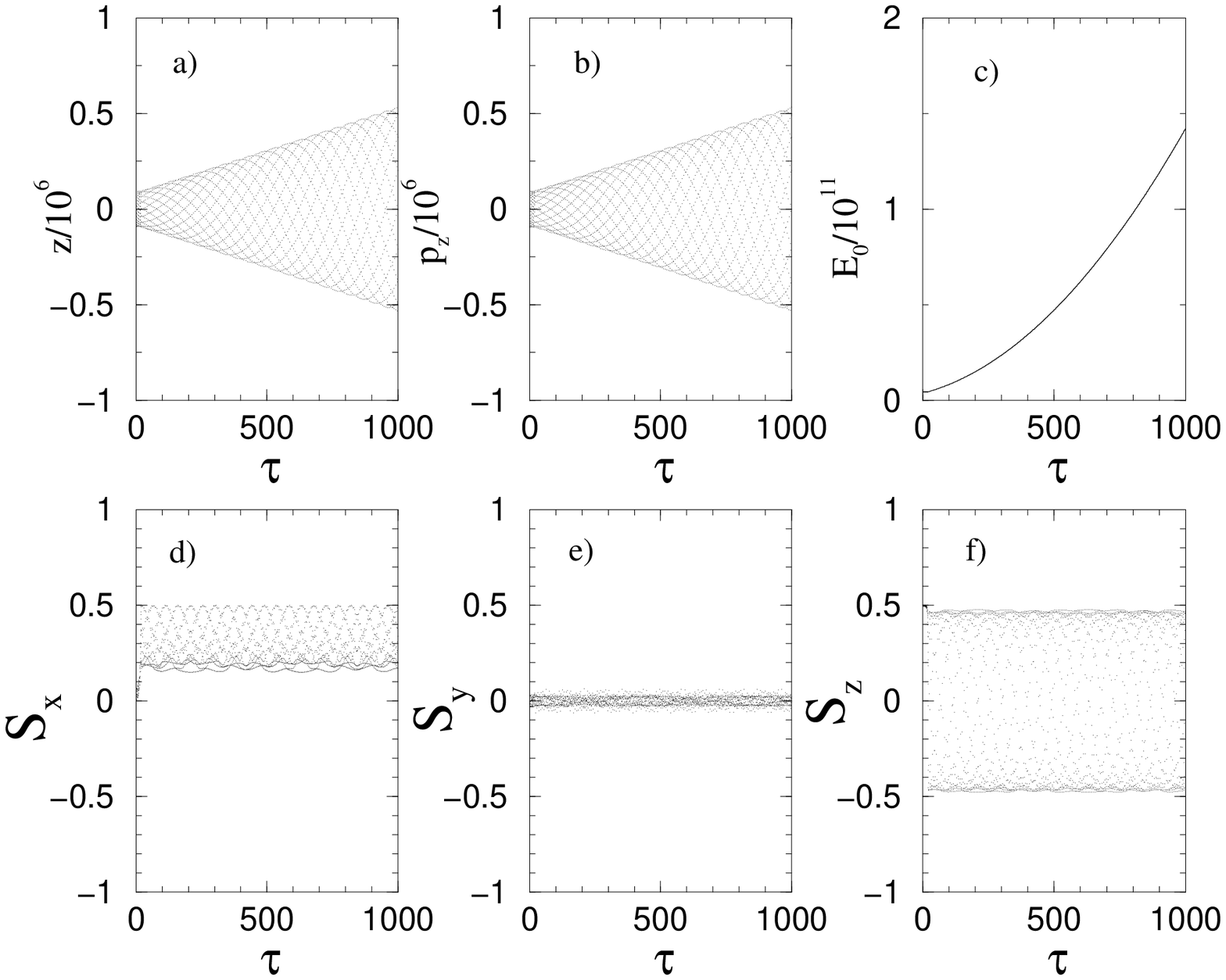} \narrowtext \caption{ Numerical
simulations of the dynamics of a classical spin-cantilever system
(equations (18)); (a)~the cantilever coordinate, $z$; (b)~ the
cantilever momentum, $ p$; (c)~the cantilever energy, $E_0 =( p^2_z+
z^2)/2$; (d,e,f)~$x,y\, \text { and}~ z $-components of the spin,
$\vec{S}$. The values of parameters are: $\varepsilon=37$, $\Delta
N=2 .9\times 10^9$, $\eta=2.8\times 10^{-7}$. The initial conditions
are: $z(0)=p_z(0)=6.7 \times 10^4 $, $ S_z( 0)=1/2$, $S_{x,y}(0)=0$.
The initial conditions, $z(0)$ and $p_z(0)$, correspond to the
root-mean-square values for $z$ and $p_z$ at room temperature. }
\label{cldyn}
\end{figure}
The corresponding dimensional value of the amplitude is, $Z\approx
24\, \text{nm}$. The experimental value in \cite{9} is $16\,
\text{nm}$, which is close to the estimated value. To estimate the
importance of nonlinear effects, we should compare the effective
``nonlinear field'', $2\eta |z|\approx 0.45$, with the transverse field,
$\varepsilon=37$. It follows that for experimental conditions
\cite{9}, the nonlinear effects are small: $2\eta
|z|/\varepsilon\approx 0.01$.

In our simulations of the classical dynamics we used the following
time dependence for $\dot\varphi$,
$$
\dot\varphi=\left\{\begin{array}{ll}
-600+30\tau, & \mbox{if $\tau \le 20$},\\
100\sin(\tau-20), & \mbox{if $\tau>20$.}
\end{array}
\right.\eqno(23)
$$
This time-dependence of $\dot\varphi$ produces cyclic adiabatic
inversion of the spin system\cite{9}. The standard condition for a
fast adiabatic inversion, $|d\vec{B}_e/dt|\ll \gamma B_1^2$, becomes:
$\ddot\varphi\ll\varepsilon^2$, which is clearly satisfied in Eqs
(23). Fig. 2 (a-f) shows the results of our numerical simulations;
these show steady growth of the cantilever vibration amplitude,
momentum and energy with time, as well as the oscillations of the
average spin, $\vec{S}$.

\section{Quantum dynamics for a single spin-cantilever system}

The magnetic force between the cantilever and a single spin is extremely small. 
To simulate the dynamics of the cantilever driven by a single spin, on
reasonable times, we take $\eta\approx 3\times 10^{-2}$. 
Such a value can be already achieved in the present day experiments\cite{9}
by measuring a single electron spin. To describe the
cantilever as a sub-system close to the classical limit, we choose the initial
wave function of the cantilever in the coherent state, $|\alpha\rangle$, in
the quasi-classical region of parameters ($|\alpha|^2\gg 1$). Namely, the
initial wave function of the cantilever was taken in the form (10), where:

$$
\Psi_1(z,0)=\sum_{n=0}^\infty A_n(0)|n\rangle,~\Psi_2(z,0)=0,\eqno(24)
$$
$$
A_n(0)=(\alpha^n/\sqrt{n!})\exp(-|\alpha|^2/2).
$$
The initial averages of $z$ and $p_z$ can be represented as,
$$
\langle z\rangle={{1}\over{\sqrt{2}}}(\alpha^*+\alpha),~\langle p_z\rangle=
{{i}\over{\sqrt{2}}}(\alpha^*-\alpha).\eqno(25)
$$

The numerical simulations of the quantum dynamics using Eqs (12)
reveal the formation of the asymmetric quasi-periodic
Schr\"odinger-cat state of the cantilever. (The dimensionless period,
$\Delta\tau=2\pi$, corresponds to the dimensional period, $\Delta
t=2\pi/\omega_c$.) Fig. 3 (a-c) shows the probability distribution,
$$
P(z,\tau)=|\Psi_1(z,\tau)|^2+|\Psi_2(z,\tau)|^2,\eqno(26)
$$
at four points in time, $\tau$. Near $\tau=40$ the probability
distribution (26) splits into two peaks; after this the separation
between these two peaks varies periodically in time. For the largest
spatial separation, the ratio of the peak amplitudes is about 100
(hence we show amplitude on a logarithmic scale).

\begin{figure}
\epsfxsize 6cm
\epsfbox{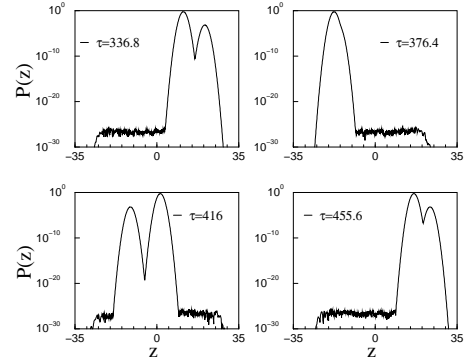}
\narrowtext
\caption{
The probability distribution for the cantilever position, $P(z)=|\Psi_1(z)|^2
+|\Psi_2(z)|^2$,
for four  instants of time.
The values of parameters are: the dimensionless Rabi frequency, $\varepsilon=
40$; the dimensionless magnetic force, $\eta=0.03$. The function,
$\dot\varphi(\tau)$, is taken in the form (23). The initial conditions: the
value of $\alpha$ in Eq. (24) corresponds to the average values, $\langle
z(0)\rangle=-20$, $\langle p_z(0)\rangle=0$. }
\label{prob2}
\end{figure}

Fig. 4 shows the probability distribution, $P(z)$, for the same
initial conditions as in Fig. 3, but we have increased $\eta$ and
$\varepsilon$ by a factor of 10: $\eta=0.3$ $\varepsilon=400$; the
cyclic adiabatic inversion parameters are:
$$
\dot\varphi=\left\{\begin{array}{ll}
-6000+300\tau, & \mbox{if $\tau \le 20$}\\
1000\sin(\tau-20), & \mbox{if $\tau>20$}
\end{array}
\right.\eqno(27)
$$
\begin{figure}
\epsfxsize 8cm
\epsfbox{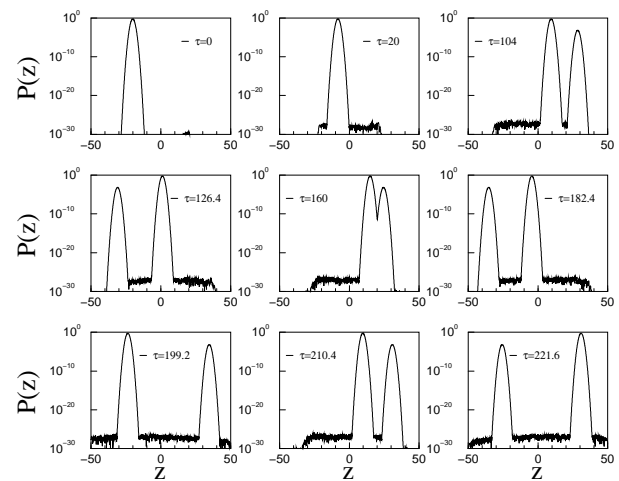}
\narrowtext
\caption{Probability distribution of the cantilever
coordinate, $z$, for $\varepsilon=400$ and $\eta=0.3$. The initial
conditions are the same as in Fig. 3.
}
\label{prob1}
\end{figure}

As shown in Fig. 4, the two peaks of the Schr\"odinger-cat state are
more clearly separated. When the probability distribution splits into
two peaks, the distance, $d$, between them initially increases. Then,
$d$ decreases. Then, the two peaks overlap and the Schr\"odinger-cat
state disappears. After this, the probability distribution splits
again so that the position of the minor peak is on the opposite side
of the major peak.  Again, the distance, $d$, first increases, then
decreases until the Schr\"odinger-cat state disappears. This cycle
repeats for as long as the simulations are run.
\begin{figure}
\epsfxsize 8cm
\epsfbox{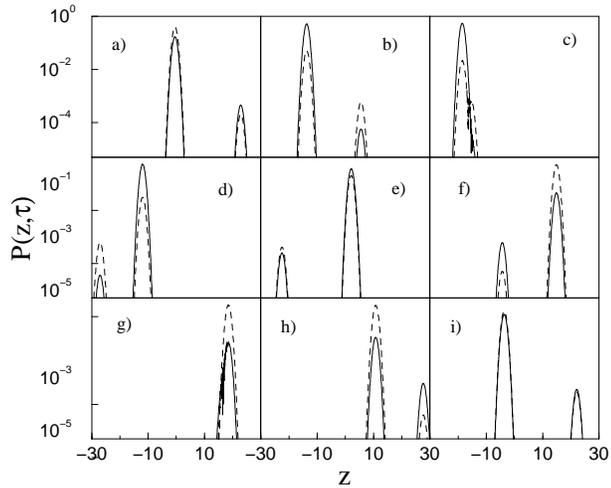}
\narrowtext
\caption{
Probability distributions,
$P_1(z,\tau)=|\Psi_1(z,\tau)|^2$ (slid curves),
 and  $P_2(z,\tau)=|\Psi_2(z,\tau)|^2$ (dashed curves)
for nine instants of time: $\tau_k=92.08+0.8k$, $k=0,1,...,8.$
}
\label{new4}
\end{figure}
\begin{figure}
\epsfxsize 8cm \epsfbox{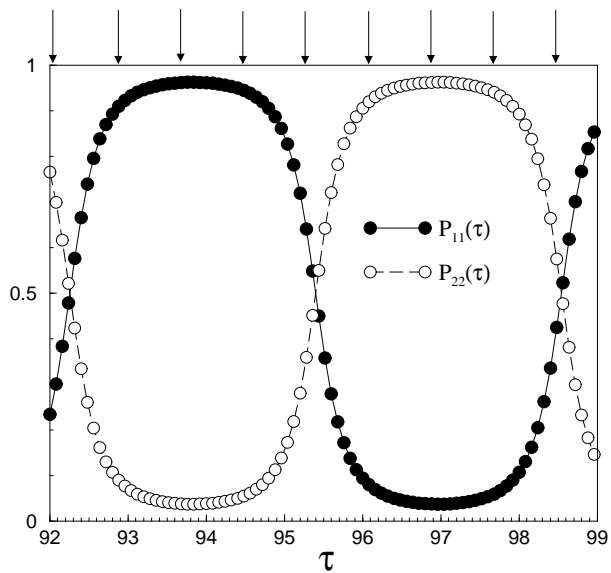} \narrowtext \caption{ Integrated
probability distributions of the spin $z$-components (diagonal components of 
the spin density matrix): $P_{11}(\tau)$,
for $S_z=1/2$ ($\bullet$); and $P_{22}(\tau)$, for $S_z=-1/2$
($\circ$), as functions of time. Vertical arrows show the time
instants, $\tau_k=92.08+0.8k$, $k=0,1,...,8$ depicted in Fig.\ 5. }
\label{newint}
\end{figure}

One might expect that the two peaks are associated with the functions
$P_n(z,\tau)=|\Psi_n(z,\tau)|^2$, $n=1,2$. In fact the situation is
more subtle: each function, $P_n(z,\tau)$ splits into two peaks. Fig.
5, shows these two functions for nine instants in time:
$\tau_k=92.08+0.8 k$, $k=0,1,...,8$. One can see the splitting of
both $P_1(z,\tau)$ and $P_2(z,\tau)$;  each peak of the function
$P_1(z,\tau)$ has the same position as the two peaks of
$P_2(z,\tau)$, but the amplitudes of these peaks differ. For instance
for $k=1$ ($\tau=92.88$) the left-hand peak is dominantly composed of
$P_1(z,\tau)$, while right hand peak is mainly composed of
$P_2(z,\tau)$. 
\begin{figure}
\epsfxsize 8cm \epsfbox{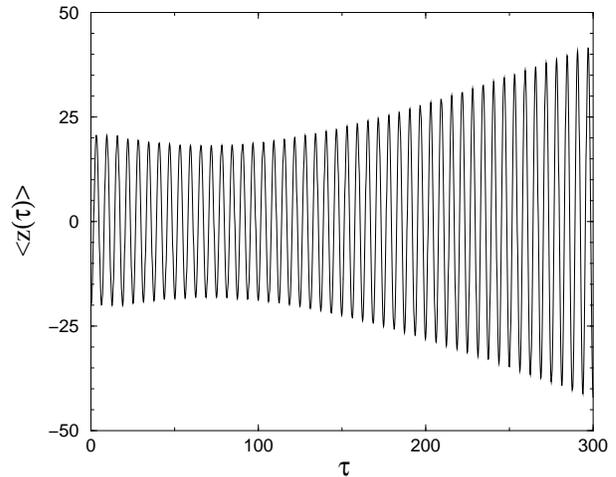} \narrowtext \caption{The dependence $\langle 
z(\tau)\rangle$ for the same values of parameters and initial conditions as
in Figs 4-6.  }
\label{newint}
\end{figure}

With increasing time the relative contribution of each
spin state to the two peaks varies.  Fig. 6 shows the relative
contribution of each spin state to the spatially integrated
probability distributions: $P_{11}(\tau)=\int P_1(z,\tau)dz$ and
$P_{22}(\tau)=\int P_2(z,\tau)dz$, as ``truly continuous'' functions of time,
$\tau$. (Vertical arrows show the time instants, $\tau_k$.) It is important 
to note that the formation of the Schr\"odinger-cat state does not suppress,
in the frameworks of the Scr\"odinger equation, the increase of the average
amplitude, $\langle z(\tau)\rangle$, of the cantilever oscillations. Fig. 7,
demonstrates the dependence $\langle z(\tau)\rangle$ for the same values of
parameters and initial conditions as in Figs 4-6.
We explored the affect of varying the initial state of the
cantilever, including the effect of an initial state which is a superposition for the spin system, and found no
significant affect on the Schr\"odinger-cat state dynamics. The value
$|\alpha|$ cannot be significantly reduced if we are going to
simulate a quasi-classical cantilever, and increasing $\alpha$
increases number of states $|n\rangle$ involved making simulation of
the quantum dynamics more difficult. Our current basis includes 2000
states which allows accurate simulations.

\section{Summary}

We consider the problem of a ``truly continuous''  measurement in
quantum physics using the example of single-spin detection with a
MRFM. Our investigation of the dynamics of a pure quantum spin 1/2 system interacting with a quasi-classical cantilever 
has revealed the generation of a quasi-periodic asymmetric
Schr\"odinger-cat state for the measurement device.

In our example of the MRFM we considered the dynamics of a
spin-cantilever system within the framework of the Schr\"odinger
equation without any artificial assumptions. For a large number of
spins, this equation describes the classical spin-cantilever
dynamics, in good agreement with previous experimental results \cite{9}. For
single-spin detection, the Schr\"odinger equation describes the
formation of a Schr\"odinger-cat state for the quasi-classical
cantilever. In a realistic situation, the interaction with
the environment may quickly destroy a Schr\"odinger-cat state
of a macroscopic oscillator (see, for example, Refs.\ \onlinecite{12,13}). 
In our case we would expect in a similar way that interaction with the environment would cause the wave function for the coupled spin-cantilever system to collapse, leading to a series of quantum jumps.

Finally, we discuss briefly the possible influence of the wave
function collapse on the quantum dynamics of the spin-cantilever
system. This collapse causes a sudden transformation of a two-peak
probability distribution into a one peak probability distribution.
It should be noted that the disappearance of the second peak does
not produce a definite value of the $z$-component of the spin because
both $P_1(z,t)$ and $P_2(z,t)$ contribute to both peaks (see Fig. 5).
Thus, after the collapse of the wave function the spin does not jump 
into one of its two stationary states, $\left|\uparrow\right\rangle$ and
$\left|\downarrow\right\rangle$, but is a linear combination 
of these states. Because of these quantum jumps, the cantilever motion (Fig. 7)
which indicates detection of a single spin may be destroyed.

As the Schr\"odinger-cat state is 
highly asymmetric (the ratio of the peak areas is of the order 100)
on average in 99 jumps out of 100 the probability distribution will
collapse into the major peak. Such jumps change the integrated
probabilities, $P_{11}$ and $P_{22}$, for the spin states with,
$S_z=1/2$ and $S_z=-1/2$. However, this change does not affect the
inequality between the values of $P_{11}$ and $ P_{22}$: If $P_{11}$
is less that $P_{22}$ (or $P_{11}$ greater than $P_{22}$) before the
collapse, the same inequality retains after the jump. On average,
only in one out of 100 jumps (when the probability distribution
collapses into the minor peak) the inequality between $P_{11}$ and
$P_{22}$ reverses. After each collapse, the system evolves according
to the Schr\"odinger equation until the Schr\"odinger-cat state
appears causing the next collapse. To simulate the dynamics of the
spin-cantilever system including quantum jumps, one
should first estimate the characteristic decoherence time, $\tau_d$,
i.e. the life-time of the Schr\"odinger-cat state taking into
consideration the interaction with the environment. Then, one can
choose a specific sequence of the life-times, $\tau_{dk}$, of the
order $\tau_d$ and a specific sequence of the wave function collapses
into the major and the minor peaks. (The probability of a collapse in
any peak is proportional to the integrated area of a peak.) After
this, one can consider quantum dynamics using Eqs (12) which
generates the Schr\"odinger-cat state and interrupted by the collapse
into one of the peaks. This rather phenomenological computational program
repeated many times with different sequences of jumps could provide
an adequate description of a possible experimental realization of a
``truly continuous'' quantum measurement which takes into account the
interaction with the environment. 

A movie demonstrating a
Schr\"odinger-cat state dynamics can be found on the WEB:\\
www.dmf.bs.unicatt.it/~$\tilde{}$ borgonov/4cats.gif

\section*{Acknowledgments}
We thank G.D. Doolen for discussions. This work  was supported by the Department of Energy under contract 
W-7405-ENG-36. The work of GPB and VIT was supported by the National Security
Agency. 
\end{document}